# Turbulence and fossil turbulence lead to life in the universe

Carl H. Gibson, Univ. Cal. San Diego,

Departments of MAE and SIO, Center for Astrophysics and Space Sciences

La Jolla, CA 92097-0411, cgibson@ucsd.edu, USA

Visiting Professor, University of Buckingham,

Buckingham Centre for Astro-Biology, Buckingham, UK

**Abstract**

Turbulence is defined as an eddy-like state of fluid motion where the inertial-vortex forces of the eddies are larger than all the other forces that tend to damp the eddies out. Fossil turbulence is a perturbation produced by turbulence that persists after the fluid ceases to be turbulent at the scale of the perturbation. Because vorticity is produced at small scales, turbulence must cascade from small scales to large, providing a consistent physical basis for Kolmogorovian universal similarity laws. Oceanic and astrophysical mixing and diffusion are dominated by fossil turbulence and fossil turbulent waves. Observations from space telescopes show turbulence and vorticity existed in the beginning of the universe and that their fossils persist. Fossils of big bang turbulence include spin and the dark matter of galaxies: clumps of ~ $10^{12}$ frozen hydrogen planets that make globular star clusters as seen by infrared and microwave space telescopes. When the planets were hot gas, they hosted the formation of life in a cosmic soup of hot-water oceans as they merged to form the first stars and chemicals. Because spontaneous life formation according to the standard cosmological model is virtually impossible, the existence of life falsifies the standard cosmological model.

## 1. Introduction

Turbulence clearly dominates mixing and diffusion in natural fluids like the ocean and atmosphere. It also controls the formation of astrophysical objects influenced by self-gravity, like stars, star clusters, galaxies and galaxy clusters, and Proto-Globular-star-Clusters of planets (PGCs). Turbulence provides the large negative stresses required by general relativity theory to drive the big bang and supply the mass-energy of inflation. Space telescopes cover an ever-widening range of frequencies, and show fossil turbulence evidence of turbulence controlling the formation and evolution of the Universe from the beginning of time to the present day.

Applications to cosmology theory of modern fluid mechanics and the revised definition of turbulence and fossil turbulence (Gibson 2012) presented in the Abstract fundamentally change the interpretation of space telescope data. Observations suggest the standard model of cosmology based on Dark Energy ($\Lambda$), Cold Dark Matter Hierarchical Clustering ($\Lambda$CDMHC), and collisionless fluid mechanics should be replaced by a new cosmology termed HydroGravitational Dynamics (HGD), Gibson (1996), Gibson & Schild (2012). The dark matter of galaxies is identified by HGD as PGC clumps of frozen primordial gas planets, as first observed and independently claimed as the missing galaxy mass by Schild (1996). Schild's interpretation and discovery of PGC planets as the dominant galaxy dark matter from quasar microlensing is further confirmed by infrared detections of the 2009 Herschel space observatory and Planck space telescope discussed below. An unanticipated result of the Gibson (1996) HGD prediction of primordial planet PGCs as the source of all stars is that this easily explains how life as observed on Earth was formed by wide and early distribution of PGCs and their planets, and water oceans formed by the planets, on cosmic scales. The controversial Hoyle-Wickramasinghe cometary panspermia hypothesis for the beginning of life on Earth, Wickramasinghe (2010), is conclusively vindicated by HGD cosmology and the new data, Gibson, Schild & Wickramasinghe (2011).

HGD cosmology rejects the underlying $\Lambda$CDMHC assumptions of collisionless, inviscid, linear, ideal, diffusionless fluid mechanics. From HGD theory, viscous stresses, turbulence, fossil turbulence, and fossil turbulence waves are critical to astrophysics and astronomy, just as they are in oceanography and atmospheric science. Summaries of the theories and observations, Gibson (2010, 2011), Gibson, Schild & Wickramasinghe (2011), are updated in the present paper. Herschel and Planck observations are discussed.

## 2. Theory

Understanding astrophysical turbulence requires that the conservation of momentum equations be applied to collisional fluids; that is, to fluids where the mean free path for collisions is smaller than the separation of fluid particles and the scale of causal connection *ct*, where *c* is the speed of light and *t* is the age of the universe since the big bang. The Navier Stokes equations are arranged so that the rate of change of specific momentum **v** equals the sum of forces per unit mass, isolating the negative gradient of the Bernoulli group B of mechanical energy terms $v^2/2 + p/\rho + lw$ (kinetic energy, enthalpy and lost work per unit mass). In most cases of interest, -grad B may be neglected. Thus the non-linear inertial vortex force term **v**×**ω** is the source of turbulence when the other forces are negligible, where the vorticity **ω** is curl **v**.

The best known criterion for turbulence to develop is the Reynolds number, which is the ratio of the inertial-vortex force **v**×**ω** to the viscous force. Boundary layers thicken until they reach a critical Reynolds number at five times the Kolmogorov length scale before they become turbulent. In stably stratified fluids the turbulence cascades to larger scales by vortex pairing and merging driven by **v**×**ω** forces. The ratio of **v**×**ω** to the buoyancy force is termed the Froude number. When this grows to a critical value, fossilization, and fossil turbulence wave radiation begins at the largest eddy sizes, Gibson, Bondur, Keeler, Leung (2011). Transport in the vertical direction in the ocean and atmosphere, and in the radial direction for self-gravitational objects, is termed fossil turbulence wave radiation. The turbulence cascade to large scales thus continues in the vertical direction until limited at a critical Froude number by buoyancy forces at the Ozmidov scale at fossilization, and in the horizontal direction until Coriolis forces cause the waves to fossilize at a critical Rossby number and Rossby radius of deformation. We will be concerned mostly with the weakly turbulent primordial plasma produced by the hot big bang as it expands and cools to form gas.

At the plasma to gas transition from HGD cosmology the Schwarz viscous fragmentation scale ~ $L_{SV} = (\gamma\nu/\rho G)^{1/2}$ rapidly decreases from that of proto-galaxies to that of proto-planets because the kinematic viscosity $\nu$ decreases by a factor of $10^{13}$, while the rate of strain $\gamma_0$ and the density $\rho_0$ maintain fossil values from the $10^{12}$ s time of first fragmentation to the $10^{13}$ s time of transition to gas (300,000 years). From heat transfer considerations, the gas also fragments at the Jeans scale $V_{Sound}/(\rho G)^{1/2}$, forming Proto-Globular-star-Cluster PGC clumps of primordial gas (H, He$^4$) planets. The size of each PGC clump of a trillion planets is $\sim (M_{PGC}/\rho_0)^{1/3}$; that is, about $3 \times 10^{17}$ meters, with planet size $\sim (M_{Earth}/\rho_0)^{1/3} = 5 \times 10^{13}$ meters. As the planets of a PGC form stars, a cavity $\sim (M_{Sun}/\rho_0)^{1/3} = 3.7 \times 10^{15}$ m is formed in the PGC clump, interpreted previously as the Oort cloud of comets, that explains the remarkable cold core filaments observed by the Planck and Herschel satellites discussed in Section 3 as proto-planetary-nebulae PPN.

In astrophysics and cosmology the most distinctive fossil of turbulence is the vorticity, conserved as angular momentum per unit area. Fossil vorticity turbulence of the big bang is preserved as weak spin anisotropies at the largest length scales of the cosmic microwave background, as shown in Figure 1 (Gibson 2012 Fig.1). A series of bumps appear in the CMB power spectrum $C_l$ of Fig. 1 that reflect vortex dynamics of big bang turbulence and plasma epoch turbulence.

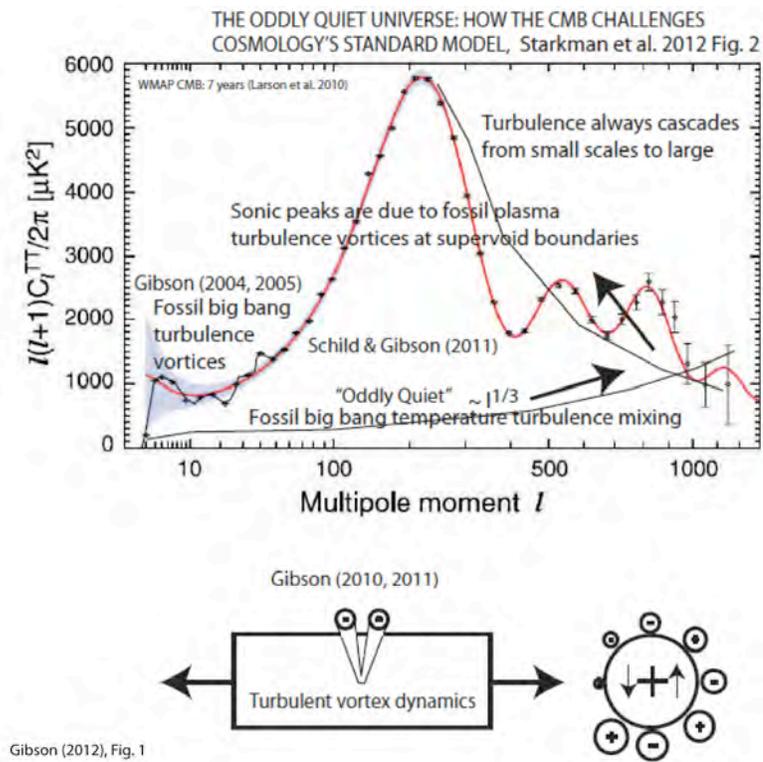

Figure 1. Cosmic microwave background temperature anisotropy spectrum from WMAP satellite.

The largest length scale bumps are on the left of Fig. 1 for $l$ = 2-40. They are labeled "Fossil big bang turbulence vortices" based on the turbulence vortex dynamics model shown at the bottom of Fig. 1 and the Gibson (2004, 2005) theory of big bang turbulence. Fossil big bang temperature turbulence should follow the indicated Corrsin-Obukhov $(1+l)l\ C_l \sim l^{1/3}$ spectral form for turbulent mixing. Cascade directions are shown by the arrows, Gibson (2012). The large amplitude bumps of Fig. 1 for wavenumbers $l > 200$ reflect secondary vortices produced by expanding superclustervoids from time $t \sim 10^{12}$ seconds after the big bang when the large kinematic viscosity of the plasma first permitted gravitational fragmentation of $10^{47}$ kg supercluster masses. Superclustervoids expand as rarefaction waves limited by the plasma sound speed $c/3^{1/2}$, where c is the speed of light. Voids of size $10^{25}$ m observed by radio telescopes are impossible by the standard ΛCDMHC cosmological model where superclusters of galaxies and the voids between them are the last objects to be formed rather than the first.

The main "sonic peak" in Fig. 1 at $l \sim 200$ reflects the size of sonic expansion reached at the time of plasma to gas transition $t \sim 10^{13}$ seconds. The fragmented supercluster objects retain the spin and density of the plasma as fossils, and so do the smaller cluster and galaxy fragments produced by further cooling before transition to gas. The fossil density from $t \sim 10^{12}$ s appears as that of globular star clusters $\rho_0 \sim 4 \times 10^{-17}$ kg m$^{-3}$. The fossil spin appears as the close alignment of rich Abell clusters of galaxies, Godlowski (2012). Because the clusters and galaxies are observed to be aligned, the standard model ΛCDMHC is falsified by Godlowski's paper. Hierarchical clustering HC of cold dark matter CDM halos would produce a random orientation of spins for rich Abell clusters. The fossil big bang turbulence spin also appears as a preferred direction on the sky termed the "axis of evil", Schild & Gibson (2011). Dipole, quadripole, etc. moments of the CMB spherical harmonic directions are found to be all pointing in the same direction; that is, along the axis of evil.

A large amount of gravitational energy remains stored in the PGC clumps of planets formed at plasma to gas transition. As shown in Figure 2, the cosmic microwave background radiation appears to be dominated by low frequency synchrotron radiation emitted by merging planets forming larger planets and the first stars soon after the plasma to gas transition, Fornengo et al. (2011).

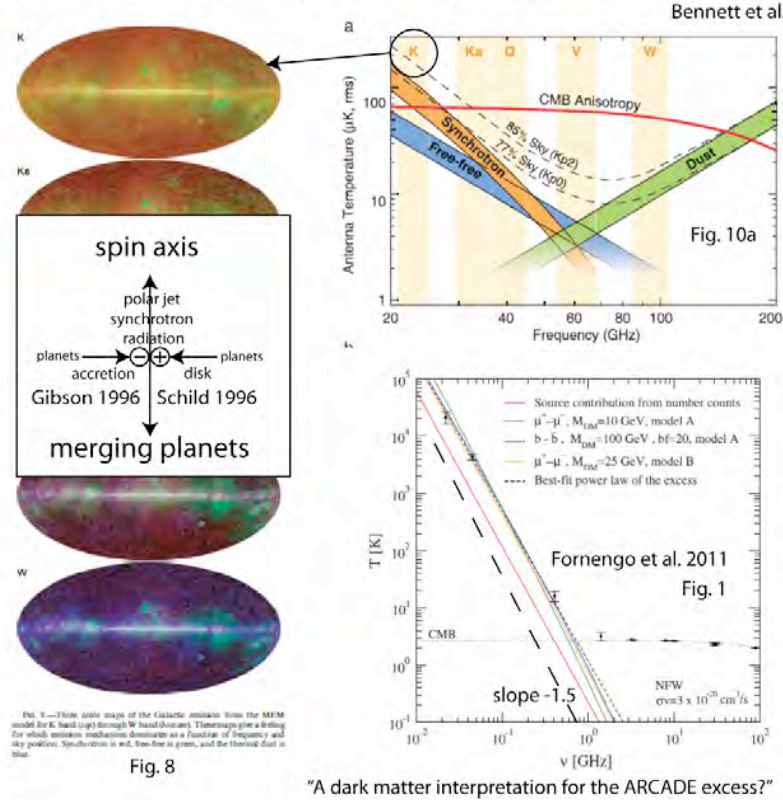

Figure 2. Low frequency synchrotron radiation (Bennett et al. 2003) from the CMB is supplied by merging dark matter planets of Gibson (1996) and Schild (1996), to explain the observed ARCADE excess (Fornengo et al. 2011).

We see from Fig. 2 (cartoon left) that polar jet synchrotron radiation from planet-merging can explain the observed CMB radiation at low frequencies. The planets merge on an accretion disk and form a plasma jet with synchrotron radiation as the central planet approaches star mass. The radiation should begin immediately after the plasma to gas transition. The first stars appear in fossil first-fragmentation gravitational free fall time $t_g = (\rho G)^{-1/2} \sim 10^{12}$ s (30,000 years) from hot gas planets at 300 thousand years, not after hundreds of millions of years of dark ages according to ΛCDMHC.

## 3. Observations

Recent observations by the Herschel space observatory support the predictions of hydrogravitational dynamics HGD cosmology. Figure 3 show an infrared image of PGCs in the Small Magellanic Cloud, with high resolution images on the left and top. A similar concentration of PGCs is found for the Large Magellanic cloud. The thousands of identical red objects detected by Herschel at 250 µ are described in the NASA webpage http://www.nasa.gov/ mission_pages /herschel/ multimedia/ pia15255.html as "dust", but are actually dark matter PGCs weakly glowing as they form larger planets.

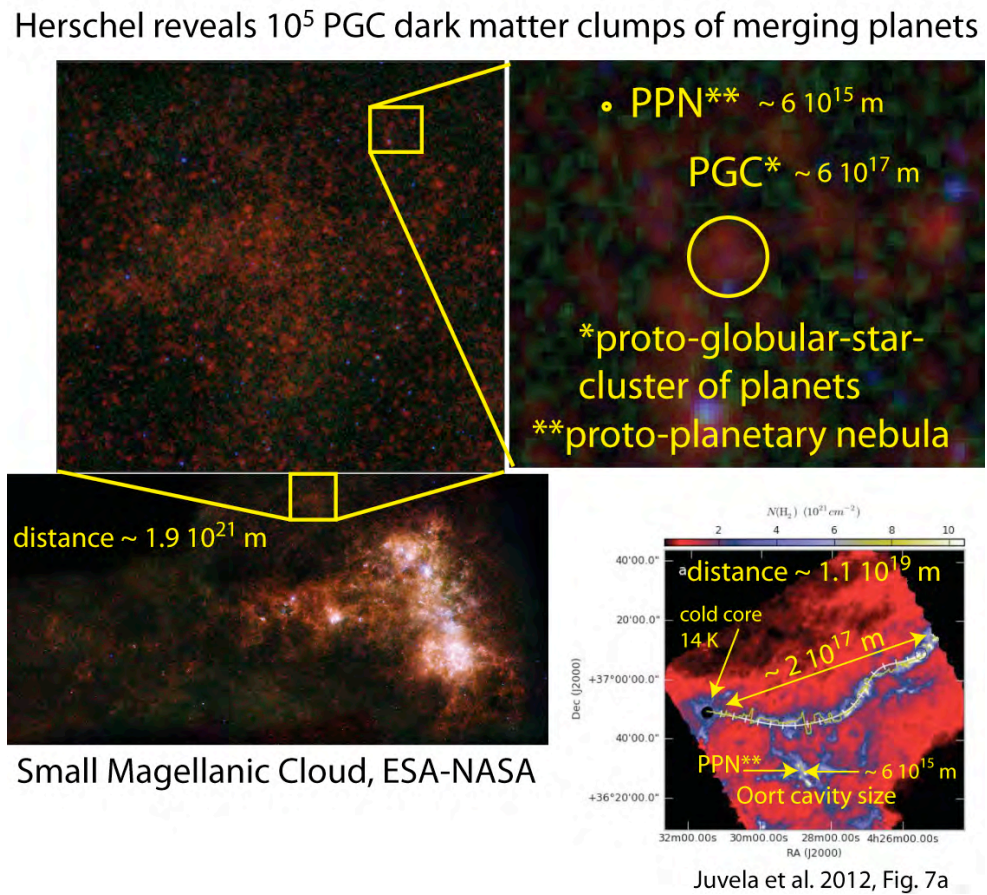

Figure 3. Herschel space observatory images of the dark matter of the Milky Way Galaxy confirms the HGD prediction that the missing mass consists of metastable PGC clumps of frozen primordial gas planets. Red dots seen in the SMC as well as the LMC star clouds are interpreted as dark matter PGCs. Image credit: ESA/NASA/JPL-Caltech/STScI

By counting the number of PGCs in the upper right hand image of Fig. 3 and measuring the area sizes in the left images it is possible to estimate ~ 100,000 PGCs in the SMC, giving a total mass of $10^{41}$ kg for the object, or about 0.1% of the mass of the Galaxy. This is close to the usually assumed value. A somewhat larger mass of PGCs is found for the Large Magellanic cloud Herschel image on the same web site. Green dots are at the Oort cavity size $6 \times 10^{15}$ m corresponding to the mass of a stellar binary, termed PPN for Proto-Planetary Nebula. According to HGD, planetary nebulae form not from material ejected by the star but from planets surrounding the Oort cavity, evaporated by polar plasma jets from central binaries such as white dwarfs nearing supernova Ia conditions due to overfeeding by cometary planets from the PGC, Gibson & Schild (2007).

Star formation in the interior of a PGC is shown by the more nearby image shown at the lower right of Fig. 3, from Juvela et al. (2012, Fig 7a). The PGC is only $1.1 \ 10^{19}$ m distant, 174 times closer than the SMC, which is at $1.9 \ 10^{21}$ m. Warm objects are detected along filaments that originate with cold core objects such as that shown on the left, with temperature 14 K matching the triple point of hydrogen. The length of the filaments are comparable to the size of a PGC, $3 \times 10^{17}$ m. The width of the filaments matches the size of an Oort cavity, $6 \times 10^{15}$ m.

The erratic positions of the filaments suggest they reflect planet and star formations induced by tidal force tracks of passing PGC centers of gravity through the observed PGC rather than the wakes of objects. It seems clear that the Herschel images of Fig. 3 are showing star formation from dark matter primordial planets in PGCs, as predicted by HGD cosmology.

## 4. Discussion

Figure 4 shows the location of the Magellanic clouds of PGCs and stars in Galactic coordinates, and their interpretation according to HGD cosmology. Protogalaxies are the smallest objects to form during the plasma epoch, just before the transition to gas at $10^{13}$ seconds. Because the PGCs are initially composed of hot primordial gas planets, they are collisional and sticky, and form large clumps such as the Magellanic clouds. The first stars form in the protogalaxy of size $L_N = 10^{20}$ meters at the Galaxy center. Most will be the small population II stars of old globular clusters, but some will be larger reflecting significant levels of turbulence, and will soon explode to form the first C, N, O, etc.

chemical oxides to seed the hot hydrogen gas planets to form water and metallic iron. At 2 million years the universe cools to the critical temperature of water 647 K so deep hot oceans form on the seeded planets. The millions of possible organic chemical reactions begin their competition for the carbon. If the complex reactions of DNA life can ever form without a miracle, this is the time. This is the biological big bang, Gibson, Wickramasinghe & Schild (2011).

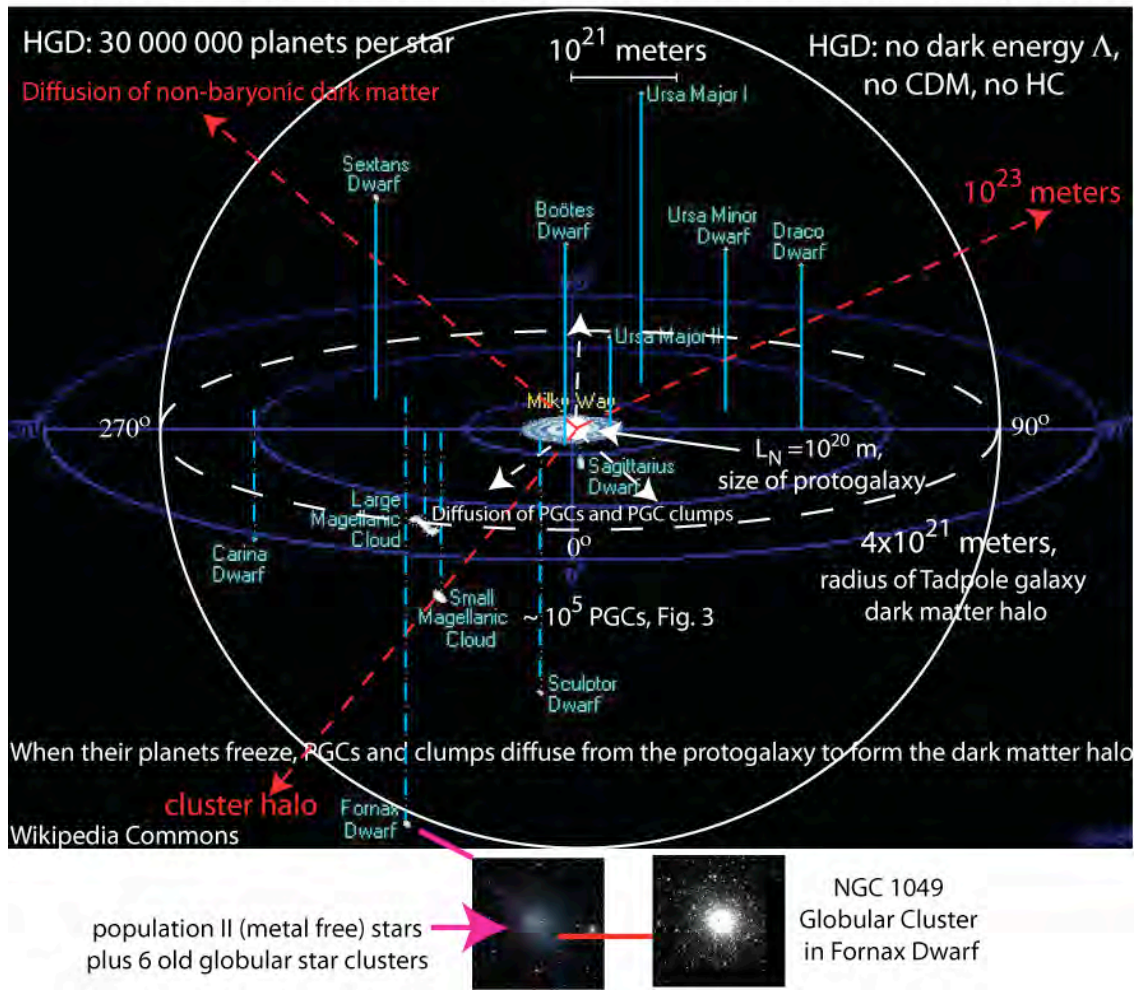

Figure 4. Herschel space observatory images of the dark matter of the Milky Way Galaxy confirms the HGD prediction that the missing mass consists of metastable PGC clumps of frozen primordial gas planets.

As shown in Fig. 4, the protogalaxy-diameter (Nomura scale) $L_N = 10^{20}$ m of the Milky Way (and all galaxies), from HGD cosmology, is 20-25 times smaller that the dark matter halo (dashed white oval) radius formed by diffusion (white dashed arrows) of the nearly collisionless PGCs resulting when their planets freeze. A sharp reduction of planet size from $10^{14}$ m at fragmentation to $10^7$ m at freezing will occur at time *t* about 30 million years when the temperature *T* of the Universe cools to the hydrogen triple point 13.8 K. The mean free path for collisions of frozen planets then becomes much larger than the PGC, so the PGCs and any clumps of PGCs that may have formed in the protogalaxy become nearly collisionless and will begin to diffuse away from the Galaxy center to form the dark matter halo. The diameter of the halo shown in Fig. 4 is the same as that observed for the Tadpole Galaxy, ~ $10^{22}$ meters. This matches the most distant of the Dwarf galaxies Fig. 4 (Fornax). An image of Fornax and one of its six globular clusters is shown at the bottom of Fig. 4. The fact that globular star clusters such as NGC 1049 are identical from galaxy to galaxy and within our Galaxy with a mass density of $\rho_0$ = $4 \times 10^{-17}$ kg m$^{-3}$ is strong evidence that the time of first fragmentation was $10^{12}$ seconds when this was the baryonic density of the expanding universe.

Life formation conditions are optimum before the diffusion of PGCs into the dark matter halo of the Milky Way in Fig. 4. Hot gas planets first formed stars, chemicals and life in the small protogalaxy. The time was 2 million years when the universe temperature decreased to the critical temperature of water 647 K so that liquid water oceans could condense to accelerate the evolution of organic chemistry. Critical temperature water is apolar and dissolves organic chemicals ordinary water will not. A cosmic soup of $10^{80}$ merging planets stirred by exploding stars and active galactic nuclei produced and distributed the complexities of DNA life to every corner of the big bang universe. The dark matter hydrogen planets cooled to the freezing point of water at 8 million years, slowing the speed of life evolution. Life formation according to ΛCDMHC cosmology cannot begin till the first star appears at hundreds of millions of years. Thus the existence of life anywhere, and on Earth, falsifies ΛCDMHC cosmology.

## 5. Conclusions

Modern fluid mechanics is needed to properly interpret the wealth of new information about cosmology provided by modern space telescopes. Gravitational structure formation starting with the big bang is much easier to understand using fluid mechanical concepts of viscosity, diffusion, turbulence, fossil turbulence and fossil turbulence waves, as employed by hydrogravitational dynamics HGD cosmology, than it is using highly questionable cold dark matter and dark energy concepts that fail to explain the space telescope observations of Figs. 1, 2 and 3. Evidence of spin alignments such as Godlowski (2012) confirms the intrinsically rotational and aligned nature of big bang turbulence vorticity fossils, Gibson (2004, 2005). Fig. 4 summarizes the application of HGD cosmology to explain the dark matter of the Milky Way as highly persistent PGC clumps of primordial planets that produce the stars and randomly dim their supernovae. Evidence that gas planets are very near all stars in all galaxies, and may produce systematic dimming errors in all supernovae 1a events, falsifies the 2011 Nobel Prize in Physics, Gibson & Schild (2011). From HGD, $\Lambda = 0$. The universe ends in a big crunch.

Perhaps the most important consequence of PGC hydrogen gas planets as the dark matter of galaxies is their crucial role in the formation of life. Because all stars form by mergers of the planets in a binary cascade from Earth mass to Solar, the organic chemistry and biological information preserved in the water oceans formed on the planets is shared on a cosmic scale starting within a few million years after the big bang. The fact that life exists on Earth rules out the possibility that the first stars and planets appeared hundreds of millions of years after the big bang, as predicted by $\Lambda$CDMHC.